\documentclass[sn-mathphys,Numbered]{sn-jnl}

\usepackage{graphicx}%
\usepackage{multirow}%
\usepackage{amsmath,amssymb,amsfonts}%
\usepackage{amsthm}%
\usepackage{mathrsfs}%
\usepackage[title]{appendix}%
\usepackage{xcolor}%
\usepackage{textcomp}%
\usepackage{manyfoot}%
\usepackage{booktabs}%
\usepackage{algorithm}%
\usepackage{algorithmicx}%
\usepackage{algpseudocode}%
\usepackage{listings}%
\usepackage{lineno, blindtext}

\theoremstyle{thmstyleone}%
\theoremstyle{thmstyletwo}%

\theoremstyle{thmstylethree}%

\begin{document}

\title[Article Title]{Superconductivity of two-dimensional hydrogenated transition-metal diborides.}


\author[1,2]{\fnm{Jakkapat} \sur{Seeyangnok}}\email{jakkapatjtp@gmail.com}

\author*[2]{\fnm{Udomsilp} \sur{Pinsook}}\email{udomsilp.p@chula.ac.th}

\author*[1]{\fnm{Graeme John} \sur{Ackland}}\email{gjackland@ed.ac.uk}

\affil[1]{\orgdiv{Centre for Science at Extreme Conditions, School of Physics and Astronomy}, \orgname{University of Edinburgh}, \state{Edinburgh}, \country{United Kingdom}}

\affil[2]{\orgdiv{Department of Physics, Faculty of Science}, \orgname{Chulalongkorn University}, \state{Bangkok}, \country{Thailand}}

    \abstract{Since the discovery of MgB$_{2}$ with T$_{c}$=39K, various metal diborides of MB$_{2}$ have been intensively studied to find possible conventional high-temperature superconductors. A possible 2D structure of the metal diboride has been shown to be in the form of M$_{2}$B$_{2}$. Using density functional theory, we investigated phase stability and possible conventional superconductors for non-hydrogenation M$_{2}$B$_{2}$, light hydrogenation M$_{2}$B$_{2}$H, and heavy hydrogenation M$_{2}$B$_{2}$H$_{4}$ of transition metal borides M$_{2}$B$_{2}$ (M=Sc,Y,V,Nb). The light hydrogenation M$_{2}$B$_{2}$H show as if they were a perturbed system from the non-hydrogenation in which the electronic structure, the phonon property, and the possible superconducting state are slightly changed. However, the heavy hydrogenation of M$_{2}$B$_{2}$H$_{4}$ give very promising 2D materials with a possible high T$_{c}$ of up to 84K at ambient pressure. This would fill the gaps for the study of possible conventional high-temperature superconductors at ambient pressure.
}

\maketitle
	\section{INTRODUCTION}
 
    The discovery of superconducting MgB$_{2}$ \cite{nagamatsu2001superconductivity} with T$_{c}$=39K has ushered in a new era in the exploration of conventional high-temperature superconductors \cite{bardeen1957microscopic}. The composition of MB$_{2}$ metal diborides has received considerable attention due to the potential for high T$_{c}$ at ambient pressure. Moreover, transition-metal diborides (MB$_{2}$) have been experimentally identified in NbB$_{2}$ \cite{leyarovska1979search}, ZrB$_{2}$ \cite{gasparov2001electron}, and TaB$_{2}$ \cite{rosner2001electronic}. These compounds possess a hexagonal crystal structure with $P6/mmm$ space-group symmetry. Following the successful fabrication of 2D MgB$_{2}$ (with T$_{c}$=36K \cite{cheng2018fabrication}), there has been a surge in research into 2D metal diborides. Recently, superconductivity in hexagonal layered transition metal diborides (MB$_{2}$) has been theoretically explored in \cite{sevik2022high}, covering M=Sc, Zr, V, Nb, Ta, Cr, and Re with T$_{c}$ values of 20.4, 2.9, 8.3, 35.5, 7.1, 4.5, and 2.4 K, respectively. Additionally, investigations have extended to superconductivity in 2D metal borides MB$_{4}$, including M=Be, Mg, Ca, Sc, and Al with T$_{c}$ values of 29.9, 22.2, 36.1, 10.4, and 30.9 K, respectively, as well as M$_{2}$B$_{2}$, where M=Mg and Re with T$_{c}$ values of 3.2 and 5.5 K, respectively.

    For bulk conventional high-temperature superconductors, attention has been primarily focused on bulk rich-hydrogen materials (hydrides) since the prediction of possible superconductivity in metallic hydrogen \cite{ashcroft2004hydrogen,mcmahon2011high}. Despite numerous successful theoretical predictions and experimental observations, such as the prediction of H$_{3}$S and LaH$_{10}$ with T$_{c}\sim$ 200 K \cite{duan2014pressure} and T$_{c}\sim$ 280 K \cite{peng2017hydrogen,liu2017potential}, and the corresponding experimental observations with T$_{c}\sim$ 203 K at 150 GPa \cite{drozdov2015conventional,einaga2016crystal} and T$_{c}\sim$ 250-260 K at 170-180 GPa \cite{drozdov2019superconductivity,somayazulu2019evidence}, respectively, these hydrides require extremely high pressure for stabilization. For more practical applications, it is crucial to explore new forms of hydrides, especially those stable at ambient pressure.  Recently, the bulk semiconductor Mg(BH$_{4}$)$_{2}$ hydride at ambient pressure has been demonstrated to exhibit metallic behavior with hole doping and a T$_{c}$ up to 140 K \cite{liu2024realizing}. Therefore, there is a growing interest in hydride forms stable at ambient pressure with potentially high T$_{c}$. Additionally, two-dimensional (2D) hydrides, particularly hydrogenated 2D materials, have undergone intensive investigation, showing promise for high superconducting temperatures.

    There are two cases of 2D hydrides. First, elements in 2D materials are substituted with hydrogen atoms, such as Janus transitional-metal dichalcogenide hydrides (JTMDs), for incomplete list \cite{seeyangnok2024superconductivity,ul2024superconductivity,li2024machine,ku2023ab,liu2022two}. Second, 2D materials are hydrogenated without removing the original atoms from pristine 2D materials, as seen in doped hydrogenated graphene with a T$_{c}$ above 90 K \cite{savini2010first}, Mo$_{2}$C$_{3}$ with a T$_{c}$ of 53 K \cite{jiao2022hydrogenation}, and CuH$_{2}$ with a T$_{c}$ greater than 40 K \cite{yan2022enhanced}. Moreover, hydrogenated MgB$_{2}$ was predicted to have a T$_{c}$ of 67 K, which could increase to 100 K when applying 5\% biaxial tensile strain \cite{bekaert2019hydrogen}, and hydrogenated phosphorus carbide, HPC$_{3}$, was also predicted to exhibit a superconducting state with a T$_{c}$ of approximately 31 K, increasing to 57.3 K when applying 3\% biaxial tensile strain \cite{li2022phonon}. Recently, two-dimensional hydrogenated metal diborides, M$_{2}$B$_{2}$H (M=Al, Mg, Mo, and W), have been predicted with T$_{c}$ values of 52.64 K, 23.25 K, 21.54 K, and 18.67 K, respectively \cite{han2023theoretical}. Furthermore, Ti$_{2}$B$_{2}$H$_{4}$ has been recently predicted to have a T$_{c}$ of 48.6 K, which can be increased to 69.4 K with a tensile strain of 9\% \cite{han2023high}.

    In this study, we systematically investigate 2D non-hydrogenated transition metal borides M$_{2}$B$_{2}$ (M = Sc, Y, V, and Nb), which have been suggested to exhibit dynamical and thermal stability in \cite{bo2018hexagonal,he2021computational}, as well as light hydrogenated transition metal borides M$_{2}$B$_{2}$H and heavy hydrogenated transition metal borides M$_{2}$B$_{2}$H$_{4}$. We begin by discussing the lattice structure, stability, electronic structure, and potential conventional high-temperature superconductivity in heavy hydrogenated transition metal borides.
	
 
    \section{RESULTS AND DISCUSSIONS.}
    \subsection{LATTICE STRUCTURE AND STABILITY.}
    \begin{figure}[h]
        \centering
		\includegraphics[width=12cm]{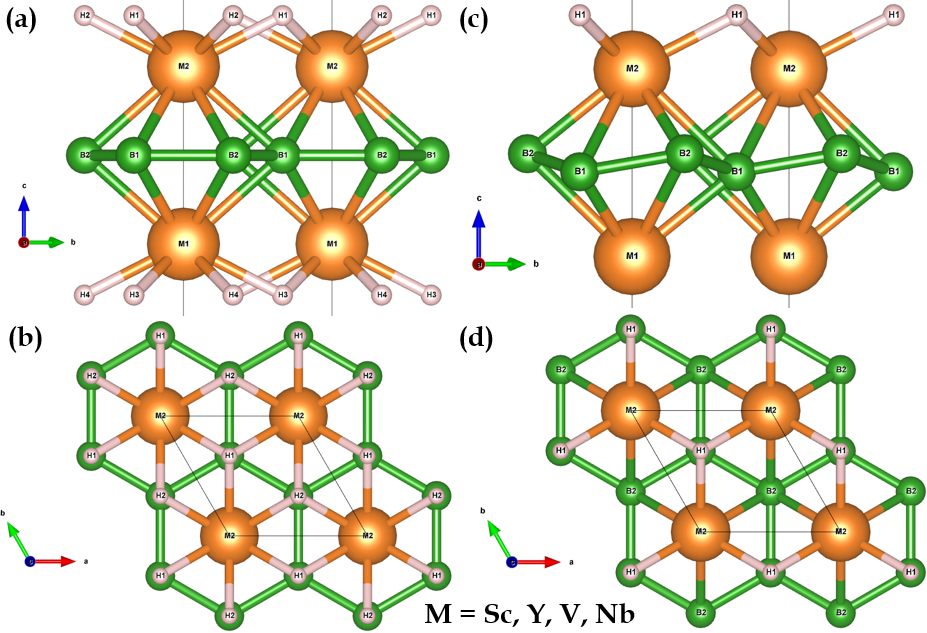}
		\caption{(a) and (b) show side and top views of the 2D M$_{2}$B$_{2}$H$_{4}$ structure (M = Sc, Y, V and Nb) where the transition metals (M), boron (B) and hydrogen (H) atoms are represented by orange, green and pink spheres, respectively. (c) and (d) show side and top views of the 2D M$_{2}$B$_{2}$H with the wrinkle of the boron layers.}
		\label{2D-m2b2h4}
    \end{figure}
    
    The crystal structure of the M$_{2}$B$_{2}$H$_{4}$ monolayer is illustrated in Fig.~\ref{2D-m2b2h4}. M$_{2}$B$_{2}$H$_{4}$ adopts a 3D hexagonal structure with a space group symmetry of $P6/mmm$, analogous to Mg$_{2}$B$_{2}$. Fig.~\ref{2D-m2b2h4} presents the side view and top view of M$_{2}$B$_{2}$H$_{x}$ (x=0,1,4) in panels (a) and (b), respectively. In this structure, boron atoms (depicted as green spheres) form a honeycomb lattice at the Wyckoff positions (1/3,2/3) and (2/3,1/2), sandwiched between two layers of transition metals M (represented by orange spheres) at the top and bottom. Hydrogen atoms (shown as pink spheres) occupy the upper and lower layers, alongside transition metals, at the Wyckoff positions (1/3,2/3) and (2/3,1/2).
    
    The optimized lattice parameters of M$_{2}$B$_{2}$H$_{4}$, M$_{2}$B$_{2}$H, and M$_{2}$B$_{2}$ are summarized in Table~\ref{lattice-parameters-table}. The lattice constants of M$_{2}$B$_{2}$H$_{4}$ and M$_{2}$B$_{2}$H (M = Sc, Y, V, and Nb) are 3.18, 3.41, 2.93, and 3.06 \AA, and 3.13, 3.32, 2.94, and 3.10 \AA, respectively. For M$_{2}$B$_{2}$, the lattice constants are 3.12, 3.29, 2.94, and 3.10 \AA, respectively.  In the case of light hydrogenated configurations, a hydrogen atom (H1) induces wrinkles in the honeycomb boron monolayer, as depicted in panels (c) and (d) of Figure~\ref{2D-m2b2h4}. The extent of wrinkling is quantified by the perpendicular distance between the two layers of wrinkled boron atoms, denoted as $\delta B$ (B2:B1), and is measured as $\delta B$ = 2.80, 3.04, 2.43, and 2.62 \AA, respectively. The distances between the two transition metal layers (M2:M1) for M$_{2}$B$_{2}$H$_{4}$ and M$_{2}$B$_{2}$H are 3.60, 3.92, 2.96, 3.32 \AA, and 3.46, 3.74, 2.83, 3.06 \AA, respectively.
    
    For the non-hydrogenated case, the distances between M2 and M1 are 3.46, 3.76, 2.80, and 3.04 \AA. The distances between the transition metals and the middle honeycomb layer of boron (M2:B1) are 1.80, 1.96, 1.48, and 1.66 \AA. In the case of non-hydrogenated and light hydrogenated configurations, these distances are 1.75, 3.12, 3.29, 2.94, and 3.10 \AA, respectively. For M$_{2}$B$_{2}$, the distances between M2 and B1 are 3.12, 3.29, 2.94, and 3.10 \AA, respectively. Furthermore, in addition to M$_{2}$B$_{2}$H$_{4}$ and M$_{2}$B$_{2}$H, the distances between the hydrogen atom layer and the boron layer (H1:B1) are 2.58, 2.78, 2.46, and 2.74 \AA for heavy hydrogenated and light hydrogenated cases, respectively.
    
    \begin{table}[h!]
        \centering
        \begin{tabular}{|c|c|c|c|c|c|c|}
        \hline
		2D materials & a (\AA) &  M2:M1 (\AA) & M2:B1 (\AA) & $\delta B$ (\AA) & H1:B1 (\AA) &$\Delta H$ (eV)\\
        \hline
        Sc$_{2}$B$_{2}$ & 3.12 & 3.46 & 1.73 & - & - & -22.9 \\
		Y$_{2}$B$_{2}$ & 3.29 & 3.76 & 1.88 & - & - & -22.3\\
		V$_{2}$B$_{2}$ & 2.94 & 2.80 & 1.40 & - & - & -28.7 \\
		Nb$_{2}$B$_{2}$ & 3.10 & 3.04 & 1.52 & - & - & -31.5\\
  		Sc$_{2}$B$_{2}$H & 3.13 & 3.46 & 1.75 & 0.040 & 2.80 & -26.7\\
		Y$_{2}$B$_{2}$H & 3.32 & 3.74 & 1.87 & 0.018 & 3.04 & -26.6\\
		V$_{2}$B$_{2}$H  & 2.94 & 2.83 & 1.46 & 0.021 & 2.43 & -32.7\\
		Nb$_{2}$B$_{2}$H & 3.10 & 3.06 & 1.56 & 0.003 & 2.62 & -35.7\\
		Sc$_{2}$B$_{2}$H$_{4}$ & 3.18 & 3.60 & 1.80 & - & 2.58 & -38.5\\
		Y$_{2}$B$_{2}$H$_{4}$  & 3.41 & 3.92 & 1.96 & - & 2.78 & -37.9\\
		V$_{2}$B$_{2}$H$_{4}$  & 2.93 & 2.96 & 1.48 & - & 2.46 & -43.1\\
		Nb$_{2}$B$_{2}$H$_{4}$ & 3.06 & 3.32 & 1.66 & - & 2.74 & -46.0\\
        \hline
	   \end{tabular}
    \caption{The table shows the lattice constants, distances between the two transition metal layers (M2:M1), distances between the transition metals and the middle honeycomb layer of boron (M2:B1), the distance between the two layers of wrinkled boron atoms $\delta B$ (B2:B1), the distances between the hydrogen atom layer and the boron layer (H1:B1), and the formation energy, respectively.}
	\label{lattice-parameters-table}
    \end{table}

    The stability of these M$_{2}$B$_{2}$H$_{4}$, M$_{2}$B$_{2}$H and M$_{2}$B$_{2}$ are explored systematically. The phonon dispersion of M$_{2}$B$_{2}$, as shown in Figure.~\ref{ph-elph-m2b2h4}, along the common high-symmetry path shows dynamical stable lattice dynamics. Furthermore, it has also recently been investigated in \cite{bo2018hexagonal,he2021computational} that Sc$_{2}$B$_{2}$, Y$_{2}$B$_{2}$, V$_{2}$B$_{2}$, and Nb$_{2}$B$_{2}$ are dynamically and thermally stable. Similarly, M$_{2}$B$_{2}$H$_{4}$ and M$_{2}$B$_{2}$H also show non-negative frequency in phonon dispersion, as shown in Fig.~\ref{ph-elph-m2b2h4}. A detailed discussion of the phonon spectrum will be discussed later. In addition, we can also examine the energy associated with the formation of a crystal structure from its constituent elements, the formation energy given by
    \begin{equation}            E_{\text{formation}}=E_{\text{structure}}-2E_{\text{M}}-2E_{\text{B}}-4E_{\text{H}},
    \end{equation}
    \begin{equation}            E_{\text{formation}}=E_{\text{structure}}-2E_{\text{M}}-2E_{\text{B}}-E_{\text{H}},
    \end{equation}
    and 
    \begin{equation}    E_{\text{formation}}=E_{\text{structure}}-2E_{\text{M}}-2E_{\text{B}},
    \end{equation}
    for M$_{2}$B$_{2}$H$_{4}$, M$_{2}$B$_{2}$H and M$_{2}$B$_{2}$, respectively where $E_{\text{M}}$, $E_{\text{B}}$ and $E_{\text{H}}$ are isolated transition metals, boron and hydrogen atom, respectively. The formation energies of M$_{2}$B$_{2}$H$_{4}$ and M$_{2}$B$_{2}$H are -38.5, -37.9, -43.1 and -46.0eV, and -26.7, -26.6, -32.7 and -35.7eV, respectively. These formation energies of 2D M$_{2}$B$_{2}$H$_{4}$ and M$_{2}$B$_{2}$H are even smaller than those of M$_{2}$B$_{2}$ where formation energies of M$_{2}$B$_{2}$ are -22.9, -22.3, -28.7, -31.5eV,respectively. These results show that the 2D M$_{2}$B$_{2}$H$_{4}$ and M$_{2}$B$_{2}$H are thermodynamic feasibility and are possible to synthesize under appropriate experimental conditions.

    \subsection{ELECTRONIC STRUCTURE.}
    \begin{figure}[h!]   
        \centering
	   \includegraphics[width=12cm]{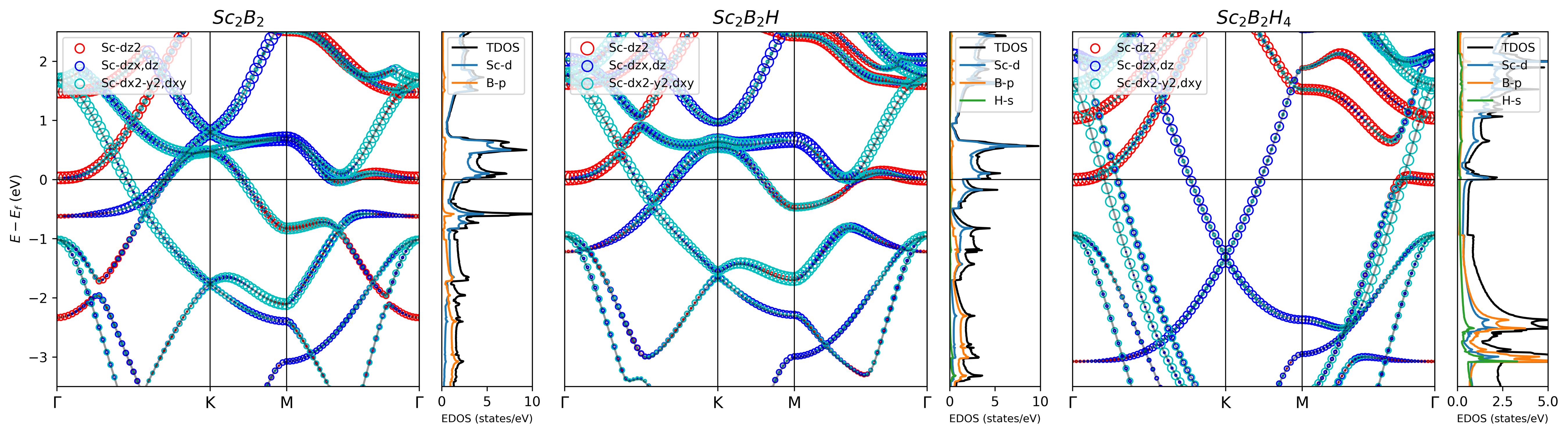}
	   \includegraphics[width=12cm]{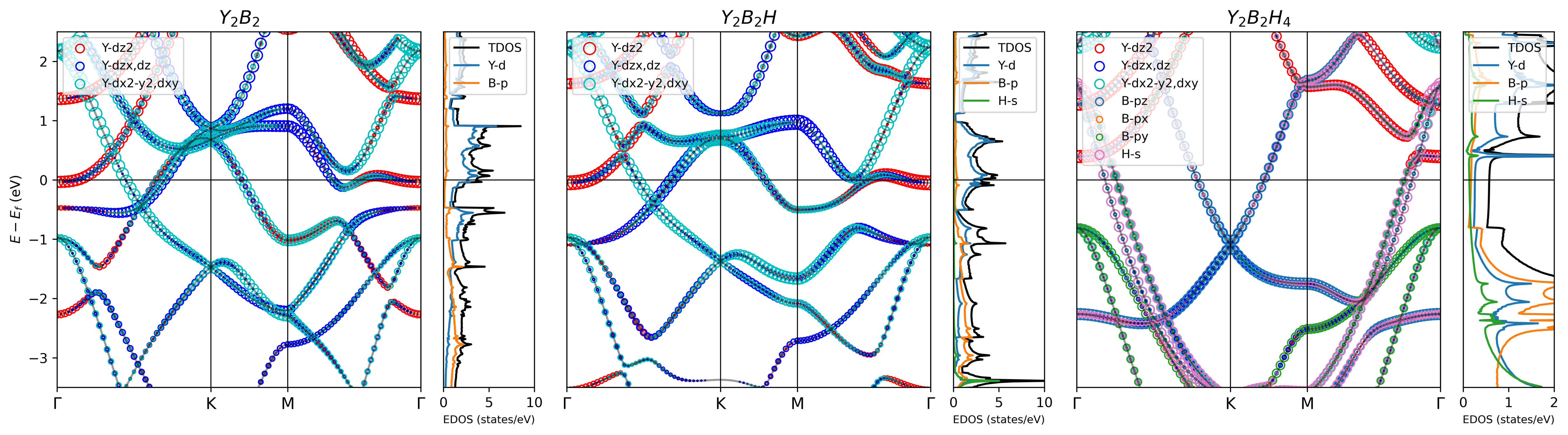}
	   \includegraphics[width=12cm]{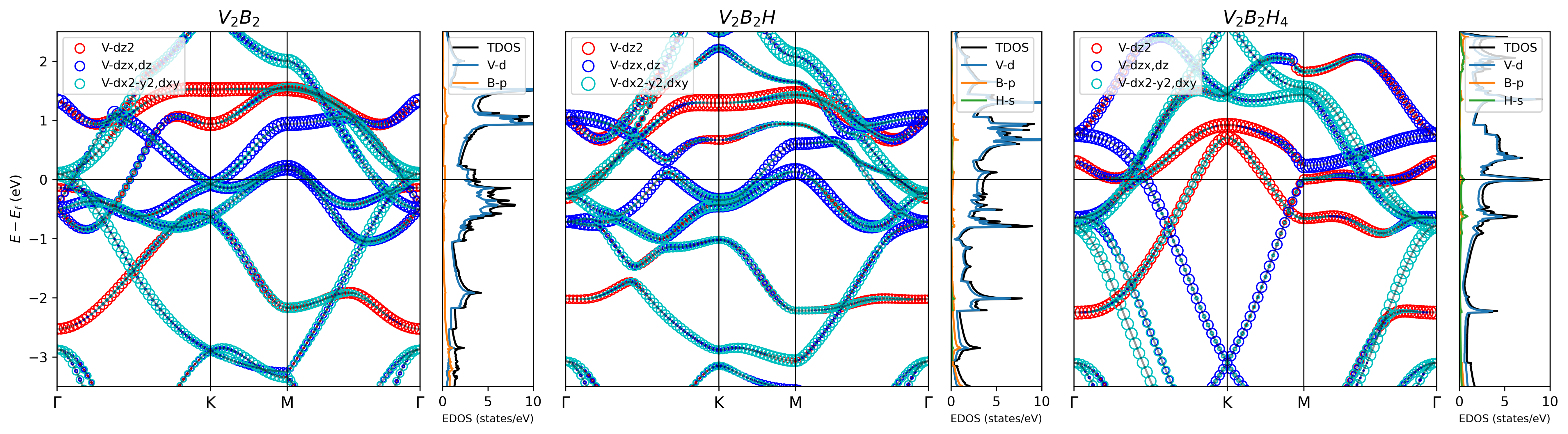}      
	   \includegraphics[width=12cm]{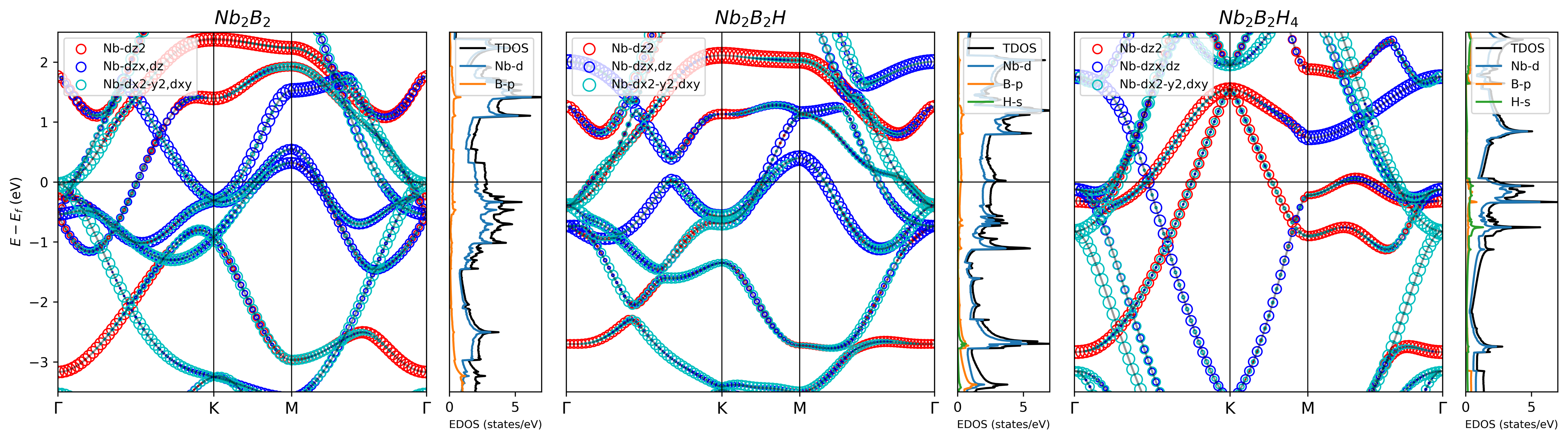}
	   \caption{The electronic structures of the 2D M$_{2}$B$_{2}$H$_{4}$, M$_{2}$B$_{2}$H, and M$_{2}$B$_{2}$ show the orbital-resolved band structure, projected electronic density of states for d-orbital of TM (Sc,Y,V,Nb), p-orbital of boron atom, and s-orbital of hydrogen atom.}
	   \label{eband-m2b2h4}
    \end{figure}
    To systematically study the electronic properties of the 2D M$_{2}$B$_{2}$H$_{4}$, M$_{2}$B$_{2}$H and M$_{2}$B$_{2}$, we will investigate the electronic structures presented in Figure~\ref{eband-m2b2h4}. In general, M$_{2}$B$_{2}$H$_{4}$, M$_{2}$B$_{2}$H and M$_{2}$B$_{2}$ show metallic behavior as a result of crossing electronic bands at the Fermi level. These crossing bands are generally dominated by the transition metal d-orbital electrons, which can be grouped into $A'(d_{z^2})$, $E'(d_{xy}, d_{x^2-y^2})$ and $E''(d_{yz}, d_{xz})$ at the $\Gamma$ point.

\begin{figure}[h]
        \centering
	   \includegraphics[width=12cm]{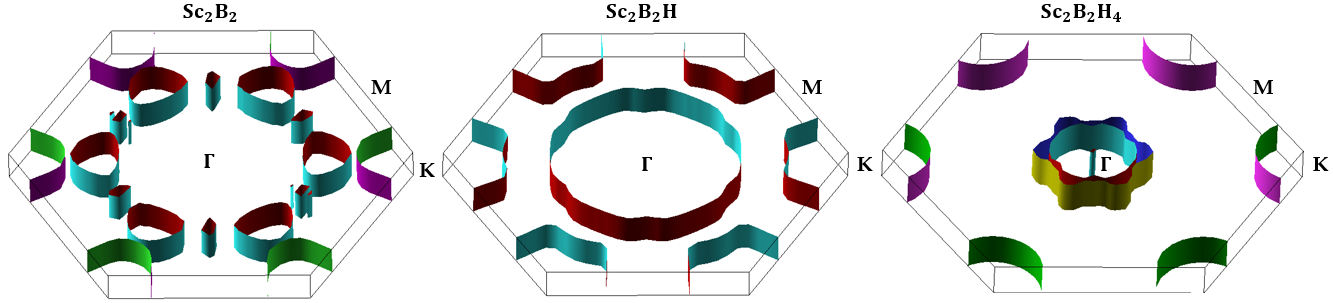}
	   \includegraphics[width=12cm]{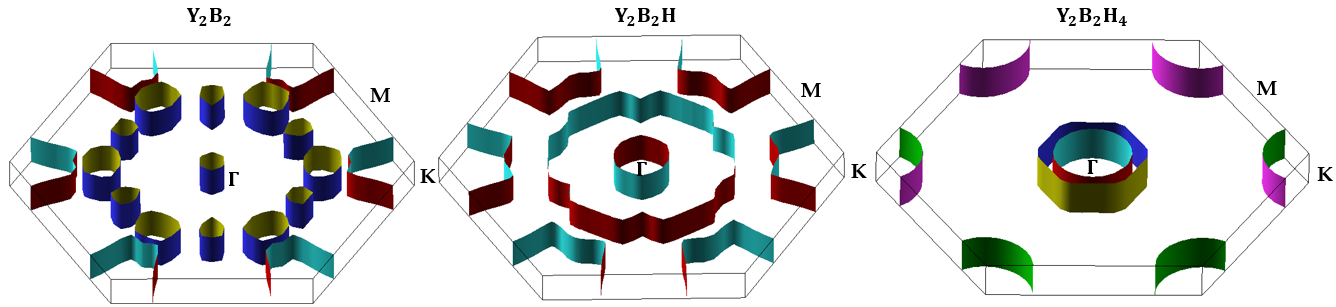}
	   \includegraphics[width=12cm]{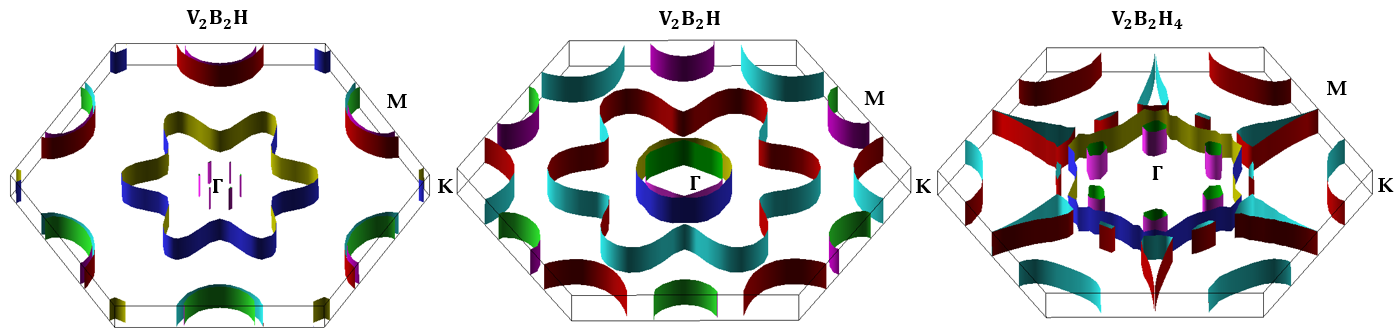}        
	   \includegraphics[width=12cm]{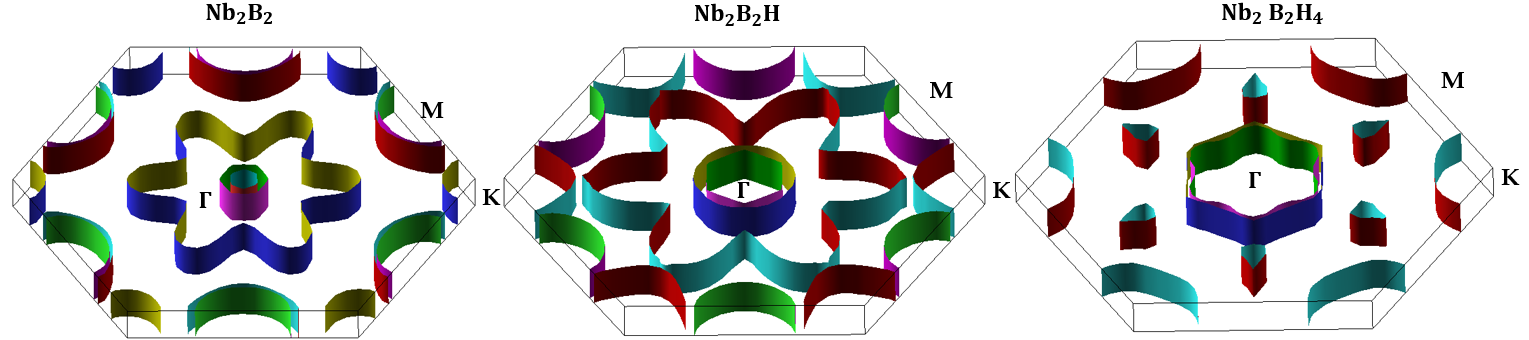}
	   \caption{These figures show the Fermi surfaces of the 2D heavy hydrogenated M$_{2}$B$_{2}$H$_{4}$, light hydrogenated M$_{2}$B$_{2}$H, and non-hydrogenated M$_{2}$B$_{2}$ where M=Sc,Y,V,Nb.}
	   \label{Fermi-m2b2h4}
    \end{figure}

    From Figure~\ref{eband-m2b2h4} showing the orbital-resolved electronic band structure, the electronic density of states (EDOS) and the orbital projected density of states (PDOS), it is clear that all pristine M$_{2}$B$_{2}$ (M=Sc,Y,V,NB) are a metal having d-orbital dominated bands crossing the Fermi level. For III-TM group (Sc,Y), Sc$_{2}$B$_{2}$ and Y$_{2}$B$_{2}$ have the similar electronic band structure, which is dominated by d-orbital electrons of $d_{z^2}$, $d_{xy}, d_{x^2-y^2}$ and $d_{yz}, d_{xz}$. For Sc$_{2}$B$_{2}$, we have three bands along the $\Gamma$ point to $K$, a single band along the $K$ point to  the $M$point, and an intersection of a single u-shaped band along the $M$ point to the $\Gamma$ point crossing the Fermi level. The intersection forms the pocket of the Fermi surface, as shown in Figure~\ref{Fermi-m2b2h4}. This also occurs for Y$_{2}$B$_{2}$ which results in the same Fermi surface topology between Sc$_{2}$B$_{2}$ and Y$_{2}$B$_{2}$. With the light hydrogenation $M_{2}$B$_{2}$H, the electronic structures are slightly modified by the additional hydrogen atom that bonds to the TM (Sc and Y). As a result, we have several Fermi surface pockets that break into a single connected Fermi surface, as shown in Figure~\ref{Fermi-m2b2h4} of Sc$_{2}$B$_{2}$H and Y$_{2}$B$_{2}$H. The electronic density of states remains almost unchanged. This indicates that the addition of a hydrogen atom slightly perturbs the electronic bands because of the small overlap between the wavefunction of the pristine system and the hydrogen atom. For the heavy hydrogenation of Sc$_{2}$B$_{2}$H$_{4}$ and Y$_{2}$B$_{2}$H$_{4}$, we have almost new electronic structures and electronic density of states, as shown in Figure~\ref{eband-m2b2h4}. The electronic density of states at the Fermi level decreases significantly as a result of the bonding of $d$-orbital electrons with additional hydrogenation. For Y$_{2}$B$_{2}$H$_{4}$, d orbital no longer totally dominates, in which the PDOS of the $p_{x}$, $p_{y}$, $p_{z}$ orbitals of the boron atom and the $s$ orbital of the hydrogen atom become comparable with d-orbital electrons from Y atoms. On the other hand, for Sc$_{2}$B$_{2}$H$_{4}$ d orbital still dominate at Fermi level, and near the $\Gamma$ point there are flat bands close to the $\Gamma$ point which results in a high electronic density of states of van Hove singularity (vHs).

    For V-TM (V, Nb) group, we have the same scenario as in the case of the III-TM group (Sc, Y) where the electronic bands and the electronic density of the state of V$_{2}$B$_{2}$H and Nb$_{2}$B$_{2}$H are slightly perturbed by an additional single hydrogen atom to the pristine compounds of V$_{2}$B$_{2}$ and Nb$_{2}$B$_{2}$. The pristine V$_{2}$B$_{2}$ have a small Fermi surface pocket, as shown in Figure~\ref{Fermi-m2b2h4}, near the $\Gamma$ point along the $\Gamma$ point to the $K$ point due to the crossing of the upper and lower bands, as shown in Figure~\ref{eband-m2b2h4}. An additional hydrogen atom slightly changes the electronic structure of V$_{2}$B$_{2}$ by removing these pockets with two shells of smooth connected Fermi surfaces near the $\Gamma$ point. For Nb$_{2}$B$_{2}$H, the electronic d-orbital dominated bands slightly shift due to the perturbation of hydrogenation, but the topology of the Fermi surface remains unchanged compared to the pristine Nb$_{2}$B$_{2}$, except that the open two-shell Fermi surfaces around the $M$ point become a single opened Fermi surface due to the lack of a crossing band along the $M$ point to $\Gamma$, as shown in Figure~\ref{Fermi-m2b2h4}.  For heavy hydrogenation, we have the special characteristics of an electronic band at the Fermi level along the $\Gamma$ point to the $M$ point that shows a flat dispersion close to the $M$ point along the $\Gamma$ point to the $M$ point, and near the $\Gamma$ point along the $\Gamma$ point to the $M$ point resulting in a high electronic density of states of van Hove singularity (vHs) at the Fermi level, as shown in Figure~\ref{eband-m2b2h4} of V$_{2}$B$_{2}$H$_{4}$. For Nb$_{2}$B$_{2}$H$_{4}$, even through we also have peaks of vHs, the Fermi level is not at any of these vHs. Therefore, this main difference between the V-TM (V,Nb) and III-TM (Sc,Y) groups is that the electronic density of the states of the III-TM (Sc,Y) group decreases significantly from non- and light hydrogenation to heavy hydrogenation, which is opposite to the V-TM group. Therefore, we would expect that the V-TM group would favor electron-phonon coupling, which we shall discuss later. These results suggest that the disrupted evolution of the electronic band topology of Fermi surfaces can be slightly changed and heavily disrupted due to the amount of extra composition elements.

    \subsection{Phonon Properties}
    
    \begin{figure}[h!]
        \includegraphics[width=13cm]{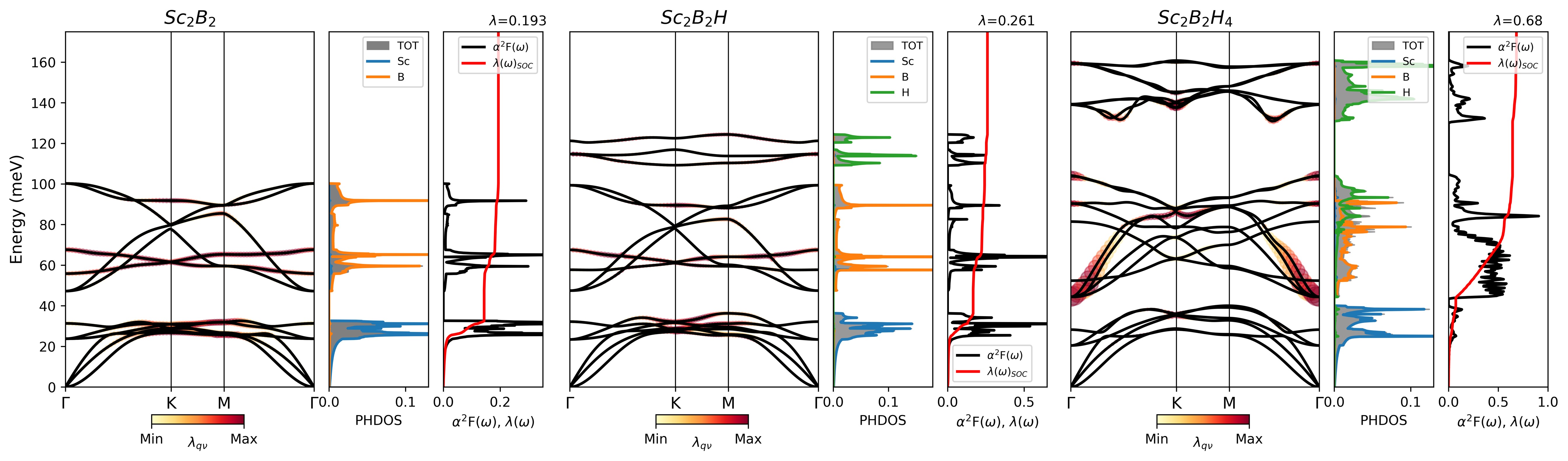}
        \includegraphics[width=13cm]{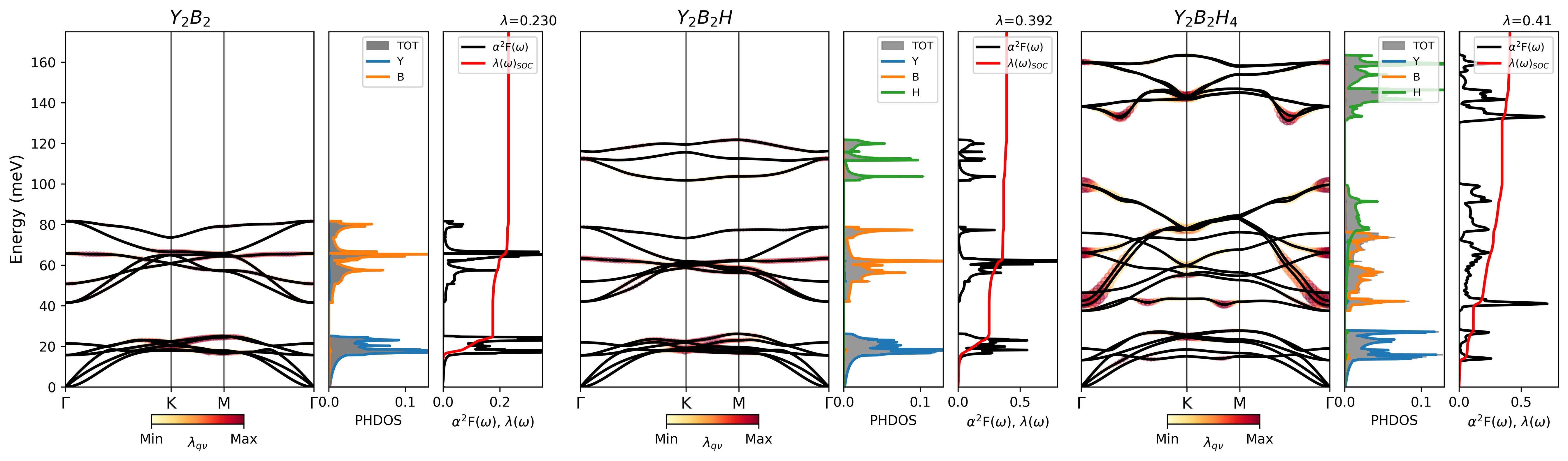}
        \includegraphics[width=13cm]{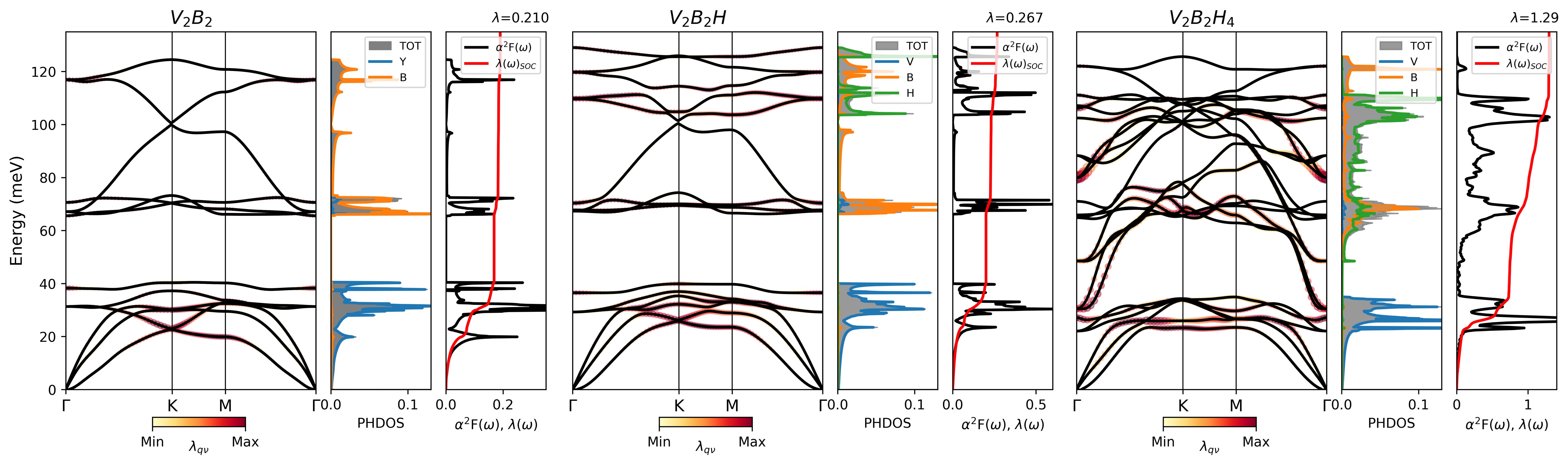}   
        \includegraphics[width=13cm]{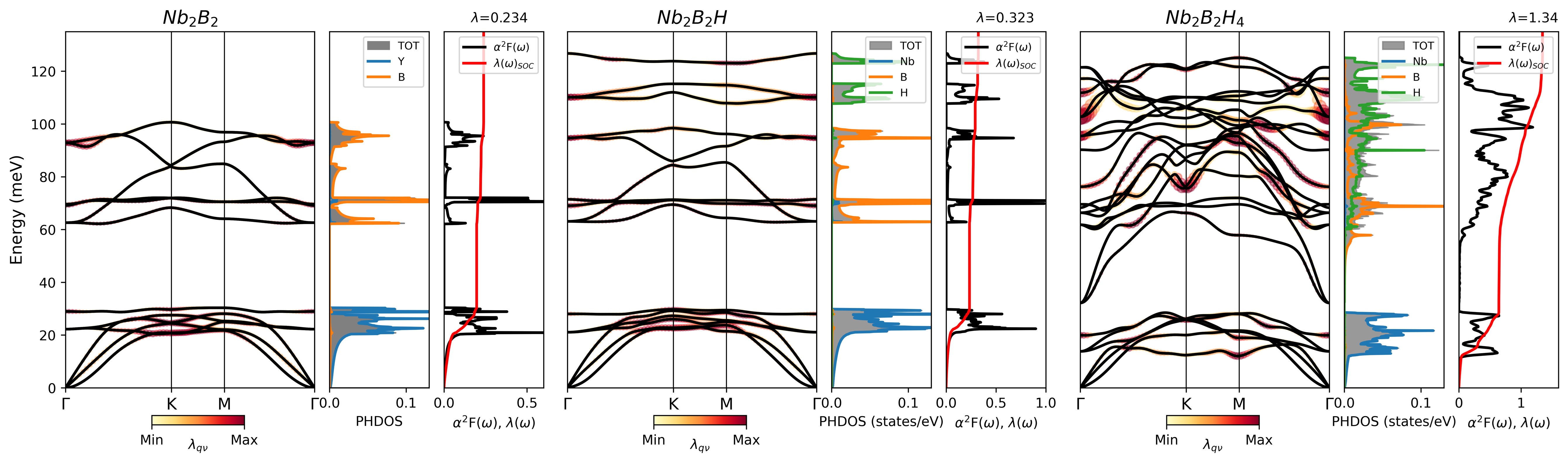}
    \caption{These plots show the weighted electron-phonon coupling (EPC) dispersion of the phonon, the projected phonon density of the states (PHDOS), the isotropic Eliashberg spectral function $\alpha^{2} F(\omega)$ and the spectrum-dependent electron-phonon coupling $\lambda (\omega)$ of 2D M$_{2}$B$_{2}$H$_{4}$, M$_{2}$B$_{2}$H, and M$_{2}$B$_{2}$.}
	\label{ph-elph-m2b2h4}
    \end{figure}
    
    As shown in Figure~\ref{ph-elph-m2b2h4}, the phonon spectrum of hydrogenated transition metal borides of M$_{2}$B$_{2}$H$_{4}$, M$_{2}$B$_{2}$H and M$_{2}$B$_{2}$ show dynamically stable lattice dynamics by having non-negative spectrum in phonon dispersion and phonon density of states. 
       
    \begin{figure}[h!]
        \centering
        \includegraphics[width=13cm]{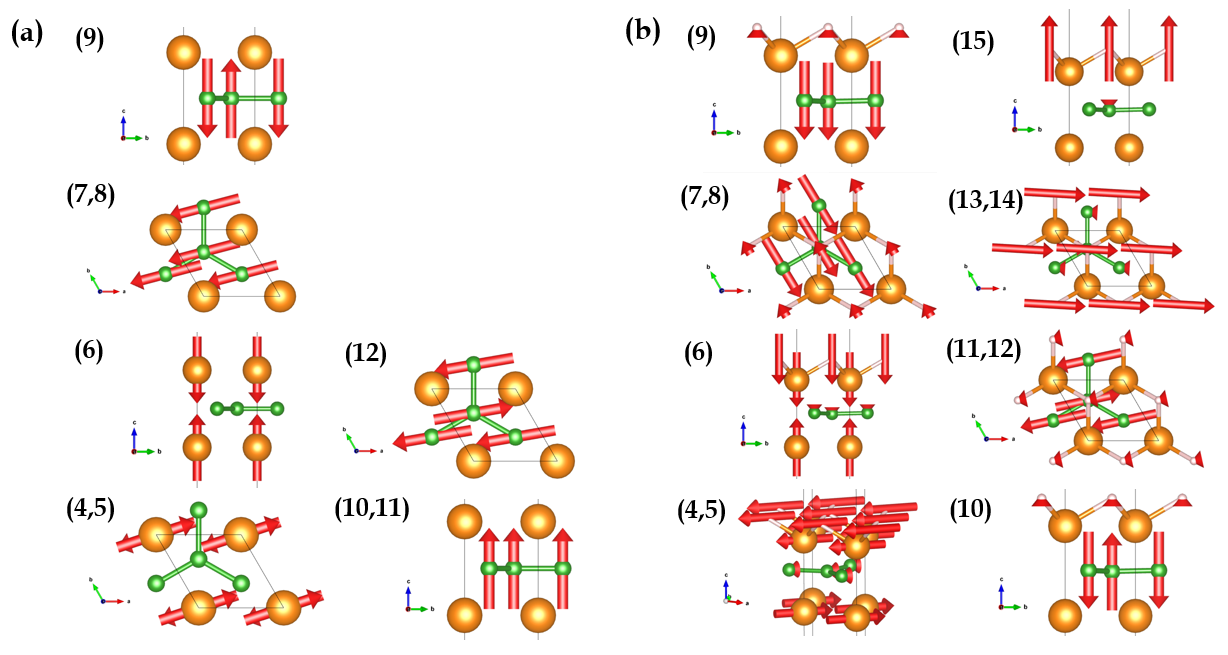}     
        \caption{ (a) shows the normal vibration modes of M$_{2}$B$_{2}$, and (b) shows the normal vibration modes of M$_{2}$B$_{2}$H (M=Sc,Y,V,Nb)}
		\label{m2b2-eigenvectors}
    \end{figure}

    For the III-TM group, we have a similar phonon dispersion between Sc$_{2}$B$_{2}$ and Y$_{2}$B$_{2}$, except that the phonon spectrum in general for Y$_{2}$B$_{2}$ is suppressed to a lower energy because of a larger mass of yttrium. Mode 9 (out-of-plane boron vibration) of Y$_{2}$B$_{2}$ also becomes a higher energy mode than mode 10 (out-of-plane boron vibration) compared to Sc$_{2}$B$_{2}$ as shown in Table~\ref{tab:eigenenergy-m2b2} for eigenvalues at the $\Gamma$ point where the eigenvectors for each modes of M$_{2}$B$_{2}$ are shown in (a) of Figure~\ref{m2b2-eigenvectors}.  For the III-TM group, we have three acoustic modes, namely a longitudinal in-plane acoustic mode (LA), a transverse acoustic in-plane mode (TA), and a flexural out-of-plane acoustic mode (ZA) that ranges from higher to lower energy, respectively, as shown in Figure~\ref{ph-elph-m2b2h4} when moving away from the $\Gamma$ point. The nine optical modes correspond to the vibration of the transition-metal and boron atoms, as shown in Figure~\ref{m2b2-eigenvectors}. The transition-metal vibrations consist of two degenerate in-plane vibrations (modes 4,5) ,and an out-of-plane vibration (mode 6). The other four non-degenerate optical modes of boron vibrations result from two degenerate in-plane vibrations (modes 7,8), two degenerate in-plane vibrations (modes 11,12), an and out-of-plane vibrations (modes 9, 10, respectively), as shown in Figure~\ref{m2b2-eigenvectors}. Similarly to the III-TM group, the phonon spectrum of the V-TM group namely, V$_{2}$B$_{2}$ and Nb$_{2}$B$_{2}$, share a common phonon dispersion and phonon density of states, except that the phonon dispersion in Nb$_{2}$B$_{2}$ have a lower energy due to a larger mass of niobium (almost twice that of vanadium). The out-of-plane boron vibration (mode 9) also gets lower energy, and in-plane mode (mode 7,8) then becomes higher than mode 9. The energy eigenvalues and eigenvectors of the normal optical modes of M$_{2}$B$_{2}$ is summarized in Table~\ref{tab:eigenenergy-m2b2}, and shown in Figure~\ref{m2b2-eigenvectors}.

    \begin{table}[h!]
    \centering
	   \begin{tabular}{|c|c|c|c|c|}
        \hline
		Modes (Eigenvectors) & $\omega$(Sc$_{2}$B$_{2}$)meV & $\omega$(Y$_{2}$B$_{2}$)meV & $\omega$(V$_{2}$B$_{2}$)meV & $\omega$(Nb$_{2}$B$_{2}$)meV \\
        \hline
		4,5 (IP) & 23.76 & 15.72 & 31.31 & 22.21 \\
		6 (OP)   & 31.33 & 21.45 & 38.24 & 28.94 \\
		7,8 (IP) & 47.32 & 41.64 & 66.76 & 62.65 \\
		9 (OP)   & 55.89 & 65.77 & 65.54 & 69.11 \\
        10 (OP) & 67.62 & 50.86 & 70.48 & 69.65 \\
        11,12 (IP) & 100.21 & 81.67 & 116.89 & 92.87 \\
        \hline
		\end{tabular}
    \caption{For the first column, it shows the eigenvectors of M$_{2}$B$_{2}$ (M=Sc,Y,V,Nb) where the direction of vibration referred as out-of-plane (OP) or in-plane (IP), and the type of atom vibrations respectively. The visualization of these eigenvectors is shown in Figure~\ref{m2b2-eigenvectors}. The other columns show the corresponding eigenvalues for each vibration modes.}
    \label{tab:eigenenergy-m2b2}
	\end{table}
 
    When hydrogenating a single hydrogen atom in the cases of Sc$_{2}$B$_{2}$H and Y$_{2}$B$_{2}$H, we have perturbed systems in which the perturbation slightly changes the original system and the vibrations of hydrogen atoms are also coupled with transition-metal atoms and boron. However, we still have the same eigenvectors of M$_{2}$B$_{2}$ that couples the vibrations of hydrogen atoms as shown in (b) of Figure~\ref{m2b2-eigenvectors} with three additional optical modes (mode 13,14,15). These three optical phonon spectrum results primarily from two degenerate in-plane boron-coupled hydrogen vibrations (mode 13,14), and an out-of-plane boron-coupled hydrogen vibration (mode 15) as the highest energy optical phonon mode in M$_{2}$B$_{2}$H. The three extra optical modes of a hydrogen atom vibration give the highest energy spectrum, obviously separated from the other nice optical modes. For Nb$_{2}$B$_{2}$H, the situation is similar for the eigenvectors and eigenvalues to that of the III-TM group, as given in Table~\ref{tab:energy-normal-modes-m2b2h}. However, the spectrum of in-plane hydrogen-coupled boron vibrations (mode 11,12) becomes higher than in-plane boron-coupled hydrogen vibrations (mode 13,14). This is why the contribution of boron vibrations becomes significant in the highest energy spectrum of V$_{2}$B$_{2}$H as shown in the projected phonon density of states Figure~\ref{ph-elph-m2b2h4}. The energy eigenvalues and eigenvectors of the normal optical modes of M$_{2}$B$_{2}$H is summarized in Table~\ref{tab:energy-normal-modes-m2b2h}, and shown in Figure~\ref{m2b2-eigenvectors}.

    \begin{table}[h!]
    \centering
	   \begin{tabular}{|c|c|c|c|c|}
        \hline
		Modes (Eigenvectors) & $\omega$(Sc$_{2}$B$_{2}$H)meV & $\omega$(Y$_{2}$B$_{2}$H)meV & $\omega$(V$_{2}$B$_{2}$H)meV & $\omega$(Nb$_{2}$B$_{2}$H)meV \\
        \hline
		4,5 (IP) & 23.40 & 15.62 & 29.79 & 19.62 \\
		6 (OP) & 32.19 & 21.97 & 36.62 & 27.35 \\
		7,8 (IP) & 47.41 & 42.06 & 67.38 & 62.75 \\
		9 (OP)  & 57.82 &  52.00 & 70.33 & 69.56 \\
        10 (OP) & 67.62 & 63.37 & 67.74 & 71.98 \\
        11,12 (IP) & 99.32 & 78.79 & 119.77 & 94.43 \\
        13,14 (IP) & 114.68 & 112.41 & 109.80 & 110.05 \\
        15 (IP) & 121.18 & 116.16 & 128.99 & 126.70 \\
        \hline
		\end{tabular}
    \caption{For the first column, it shows the eigenvectors of M$_{2}$B$_{2}$H (M=Sc,Y,V,Nb) where the direction of vibration referred as out-of-plane (OP) or in-plane (IP), and the type of coupled-atom (X) atom (Y) vibrations (XY), respectively. The visualization of these eigenvectors is shown in Figure~\ref{m2b2-eigenvectors}. The other columns show the corresponding eigenvalues for each vibration modes.}
    \label{tab:energy-normal-modes-m2b2h}
	\end{table}

    \begin{figure}[h!]
        \centering
        \includegraphics[width=13cm]{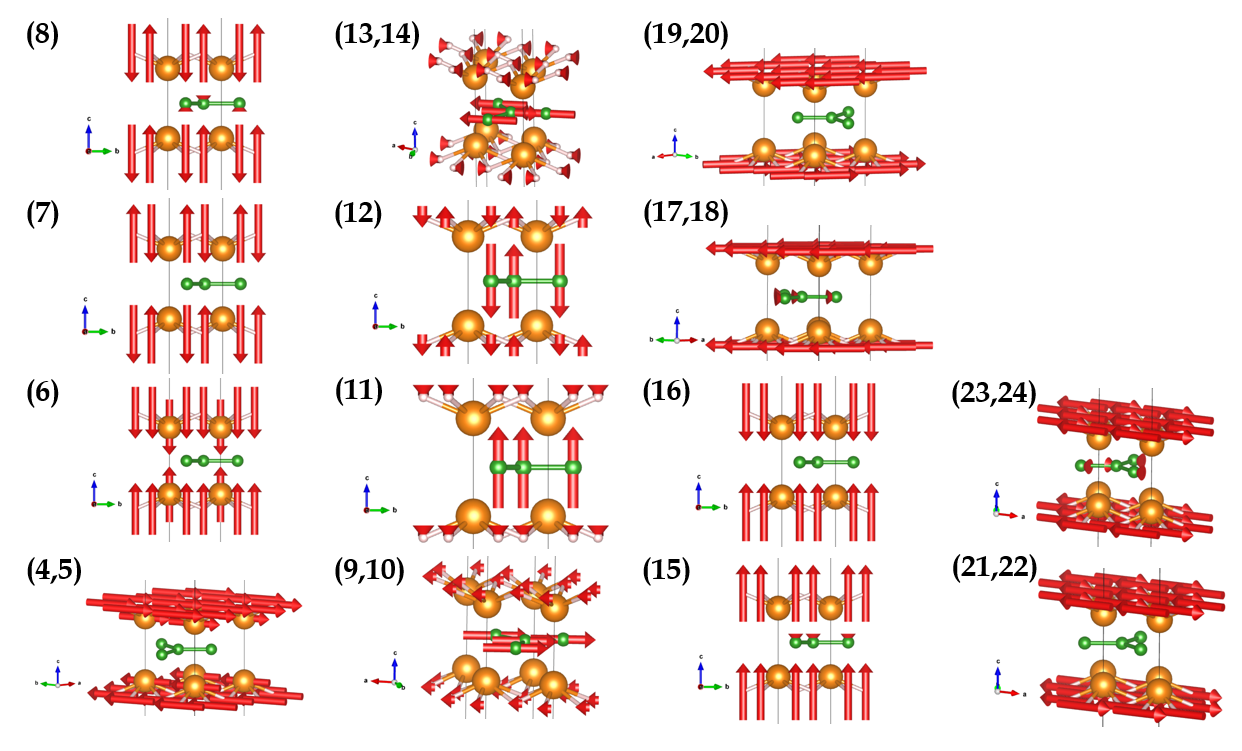} 
        \caption{shows the normal vibration modes M$_{2}$B$_{2}$H$_{4}$ (M=Sc,Y,V,Nb).}
		\label{m2b2h4-eigenvectors}
    \end{figure}

    \begin{table}[h!]
        \centering
        \begin{tabular}{|c|c|c|c|c|c|}
        \hline
		Modes (Eigenvectors) & $\omega$(Sc$_{2}$B$_{2}$H$_{4}$)meV & $\omega$(Y$_{2}$B$_{2}$H$_{4}$)meV & $\omega$(V$_{2}$B$_{2}$H$_{4}$)meV & $\omega$(Nb$_{2}$B$_{2}$H$_{4}$)meV \\
        \hline
		4,5 (IP) & 20.48  & 13.25  & 23.31  & 12.34 \\
		6  (OP)   & 28.21  & 18.97  & 28.00  & 21.34  \\
		7  (OP)   & 44.37  & 40.28  & 102.43 & 76.17\\
    	8   (OP)  & 44.40  & 42.47  & 107.11 & 90.22\\
        9,10 (IP) & 44.59  & 37.52  & 64.92  & 61.84\\
        11 (OP) & 52.44  & 46.41  & 65.93  & 68.38 \\
        12  (OP)  & 81.51  & 75.88  & 70.99  & 69.60\\
        13,14 (IP) & 90.42  & 66.13  & 122.03 & 95.46\\
        15  (OP)  & 103.75 & 99.61  & 111.07 & 117.15\\
        16  (OP)  & 104.08 & 99.51  & 106.17 & 111.80\\
        17,18 (IP) & 139.02 & 138.00 & 30.86  & 32.21\\
        19,20 (IP) & 139.31 & 138.32 & 48.78  & 32.22\\
        21,22 (IP) & 159.29 & 159.81 & 88.43  & 121.46\\
        23,24 (IP) & 159.43 & 160.04 & 80.10  & 102.73\\
        \hline
		\end{tabular}
    \caption{For the first column, it shows the eigenvectors of M$_{2}$B$_{2}$H$_{4}$ (M=Sc,Y,V,Nb) where the direction of vibration referred as out-of-plane (OP) or in-plane (IP), and the type of coupled-atom (X) atom (Y) vibrations (XY), respectively. The visualization of these eigenvectors is shown in Figure~\ref{m2b2h4-eigenvectors}. The other columns show the corresponding eigenvalues for each vibration modes.}
    \label{table:normalmodes-m2b2h4}
	\end{table}
 
     For heavy hydrogenation, the addition of four additional hydrogen atoms completely modify the phonon dispersion of non-hydrogenated cases. For the heavy hydrogenation of Sc$_{2}$B$_{2}$H$_{4}$, the phonon of in-plane vibrations from boron-coupled hydrogen vibrations and hydrogen vibrations dominates the spectrum at 140 and 160meV. Phonon at 140meV results from two degenerate in-plane boron-coupled hydrogen vibrations (modes 17,18) and two degenerate in-plane hydrogen vibrations (mode 19,20). The phonon at 160meV results from two degenerate in-plane hydrogen vibrations (mode 21,22) and two degenerate in-plane boron-coupled hydrogen vibrations (mode 23,24). This forms the first group of phonon spectrum at the highest energy.  The second group of the phonon spectrum in the moderate energy range of Sc$_{2}$B$_{2}$H$_{4}$ results from eight eigenvalues of about 40meV to 100meV. It consists of out-of-plane hydrogen vibration (mode 7), out-of-plane boron-coupled hydrogen (mode 8), two degenerate in-plane hydrogen-coupled boron vibrations (mode 9,10), out-of-plane hydrogen-coupled boron vibration (mode 11), out-of-plane hydrogen-coupled boron vibration (mode 12), two degenerate in-plane hydrogen-coupled boron vibrations (mode 13,14), out-of-plane boron-coupled hydrogen vibration (mode 15), and out-of-plane hydrogen vibration (mode 16), as shown in Figure~\ref{m2b2h4-eigenvectors}. In the lowest group of phonon spectrum, we have three acoustic modes, two degenerate in-plane transition-metal atom vibrations (mode 4,5), and an out-of-plane transition-metal atom vibration (mode 6). This forms the phonon spectrum at low energy. For Y$_{2}$B$_{2}$H$_{4}$, even through we have the shift up and down of energy eigenvalues as shown in Table~\ref{table:normalmodes-m2b2h4}. However, these shifts still occur in the same range for each group of the phonon spectrum as previously discussed for Sc$_{2}$B$_{2}$H$_{4}$. As a result, the phonon spectrum of Sc$_{2}$B$_{2}$H$_{4}$ and Y$_{2}$B$_{2}$H$_{4}$ is occupied three regions of energies, as shown in Figure~\ref{ph-elph-m2b2h4}. 
     
     For the V-TM group, the phonon dispersion is also completely modified by hydrogenation, similar to that for the group III-TM. The main difference is that the phonon spectrum of hydrogen and boron-coupled hydrogen vibrations have lower energies. On the other hand, the phonon spectrum of hydrogen-coupled boron vibrations has a higher energy, as summarized in Table~\ref{table:normalmodes-m2b2h4} for the detailed shift of each eigenvalue of the vibration. Therefore, the phonon spectrum of V$_{2}$B$_{2}$H$_{4}$ and Nb$_{2}$B$_{2}$H$_{4}$ is occupied two regions of energies, as shown in Figure~\ref{ph-elph-m2b2h4}. In the case of V$_{2}$B$_{2}$H$_{4}$, the highest phonon energy no longer results from hydrogen vibration (mode 21,22 for Nb$_{2}$B$_{2}$H$_{4}$) or boron-coupled hydrogen vibration (mode 23,24 for Sc$_{2}$B$_{2}$H$_{4}$ and Y$_{2}$B$_{2}$H$_{4}$) as shown in Table~\ref{table:normalmodes-m2b2h4}, rather than two degenerate hydrogen-coupled boron vibrations (Mode 12,13). This is the reason why we have boron as the main contribution to the phonon spectrum, as shown in the projected phonon density of the states of V$_{2}$B$_{2}$H$_{4}$ of Figure~\ref{ph-elph-m2b2h4}.
     
    \begin{figure}[h]
        \centering
        \includegraphics[width=12cm]{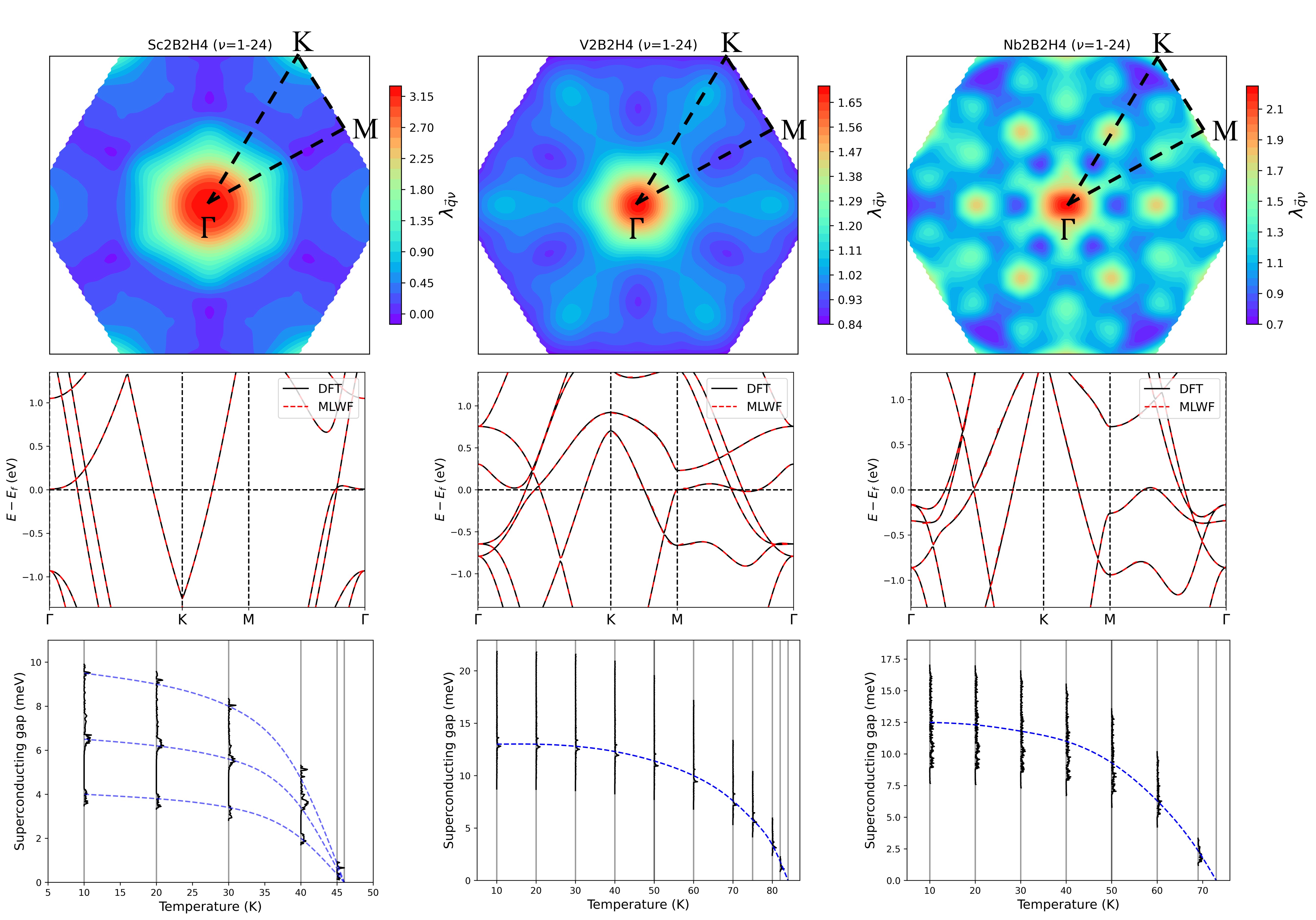}   
        \caption{The three-upper images show the electron-phonon coupling distributed over the Brillouin zone, the three-lower middle images show the comparison between the electronic bands obtained from DFT and maximally-localised Wannier function, and the three-lower images show the superconducting gap of Sc$_{2}$B$_{2}$H$_{4}$, V$_{2}$B$_{2}$H$_{4}$, and Nb$_{2}$B$_{2}$H$_{4}$, respectively.}
		\label{fig:epc-gaps}
    \end{figure}    
     
    \subsection{Electron-Phonon coupling and phonon-mediated superconductivity.}

    The electron-phonon coupling (EPC) is investigated via the weighted EPC dispersion of the phonon and the contour plot of EPC ($\lambda_{qv}$) as shown in Fig.~\ref{ph-elph-m2b2h4} and Fig.~\ref{fig:epc-gaps}, respectively. In general, the EPC primarily arises near the $\Gamma$ point for Sc$_{2}$B$_{2}$H$_{4}$, V$_{2}$B$_{2}$H$_{4}$, and Nb$_{2}$B$_{2}$H$_{4}$. For Sc$_{2}$B$_{2}$H$_{4}$, this contribution comes from the optical vibration modes $\nu = 7-10$, specifically the out-of-plane hydrogen vibrations ($\nu = 7, 8$) and the out-of-plane boron vibrations ($\nu = 9, 10$). For V$_{2}$B$_{2}$H$_{4}$, the EPC is largely contributed near the $\Gamma$ point ($\nu = 6-8, 15-16$), but it also results from various vibration modes away from the $\Gamma$ point. This behavior is observed in the weighted EPC dispersion of the phonon in the energy range around 70 meV between the $K$ point and the $M$ point, as shown in Fig.~\ref{ph-elph-m2b2h4}. This leads to a high average electron-phonon coupling function of the Eliashberg function, with significant contributions around 30 meV, 70 meV, and 100 meV, resulting in a large $\lambda$ value of 1.29. For Nb$_{2}$B$_{2}$H$_{4}$, the Eliashberg function is divided into two regions: lower energy and higher energy. The EPC at lower energy, around 20 meV, results from the phonon excitation of niobium away from the $\Gamma$ point. At higher energy, the EPC is distributed throughout the Brillouin zone for the optical vibrations of boron and hydrogen. Consequently, the net contour map of EPC shows peaks of $\lambda_{qv}$ throughout the Brillouin zone, leading to the highest value of $\lambda$.

    \begin{table}[h!]
    \centering
	   \begin{tabular}{|c|c|c|c|c|c|c|}
        \hline
		2D materials & $\lambda$ & $\omega_{\text{log}}$ (meV) & $\omega_{2}$ (meV) & T$_{c}$ (K) [AD] & T$_{c}$ (K) [ME] & Refs. \\
        \hline
        Al$_{2}$B$_{2}$ & 0.29 & 21.05 & - & 0.10 &- & \cite{han2023theoretical} \\
        Mg$_{2}$B$_{2}$ & 0.43 & 42.41 & - & 3.26 &- & \cite{han2023theoretical,sevik2022high} \\
        Mo$_{2}$B$_{2}$ & 0.40 & 25.04 & - & 1.12 &- & \cite{han2023theoretical} \\
        W$_{2}$B$_{2}$ & 0.38 & 19.90 & - & 0.31 &- & \cite{han2023theoretical} \\
        Re$_{2}$B$_{2}$ & 1.088 & 6.966 & - & 5.5 &- & \cite{sevik2022high} \\
        Ti$_{2}$B$_{2}$ & 0.31 & 30.16 & - & 0.0 &- & \cite{han2023high} \\
        Ti$_{2}$B$_{2}$H$_{4}$ & 1.18 & 47.35 & - & 48.6 &- & \cite{han2023high} \\
        Sc$_{2}$B$_{2}$ & 0.19 & 33.6 & - & 0.0 &- & This work \\
		Y$_{2}$B$_{2}$  & 0.23 & 29.3 & - & 0.0 &- & This work \\
		V$_{2}$B$_{2}$ & 0.21 & 26.6 & - & 0.0 &- & This work \\
		Nb$_{2}$B$_{2}$ & 0.23 & 24.7 & - & 0.0 &- & This work \\
  		Sc$_{2}$B$_{2}$H & 0.26 & 38.7 &- & 0.0 &- & This work \\
		Y$_{2}$B$_{2}$H  & 0.39 & 29.5 & - & 1.3 &- & This work \\
		V$_{2}$B$_{2}$H  & 0.27 & 34.8 & - & 0.0 &- & This work \\
		Nb$_{2}$B$_{2}$H & 0.32 & 31.9 & - & 0.4 &- & This work \\
		Sc$_{2}$B$_{2}$H$_{4}$ & 0.68 & 49.26 & - & 18.73 & 46 & This work \\  
		Y$_{2}$B$_{2}$H$_{4}$  & 0.41 & 49.52 & - & 2.74 &- & This work \\
		V$_{2}$B$_{2}$H$_{4}$  & 1.29 & 42.33 & 58.86 & 53.35 & 84 & This work \\
		Nb$_{2}$B$_{2}$H$_{4}$ & 1.34 & 40.88 & 67.14 & 54.40 & 73 & This work \\
        \hline
	   \end{tabular}
    \caption{The table shows the superconducting quantities including the electron-phonon coupling constant $\lambda$, the logarithmic average of the phonon energy $\omega_{\text{log}}$, the square average of the phonon energy $\omega_{2}$, the critical superconducting temperature T$_{c}$ [AD] based on the Allen-Dynes formula, and the critical superconducting temperature T$_{c}$ [ME] based on the closing of the superconducting gap.}
	\label{superconducting-quantities-table}
	\end{table}

    According to Eq.(\ref{eq:eliashberg}), the Eliashberg spectral function depends on the product of the transition matrix, the Fermi nesting function, and the phonon spectrum. To obtain a high superconducting temperature (T$_{c}$), we need large values of electron-phonon coupling ($\lambda$), and a high logarithmic average of the phonon spectrum $\omega_{\text{log}}$. Important values to determine the superconducting temperature are shown in Table~\ref{superconducting-quantities-table} together with other possible superconductors that have previously been studied. For non-hydrogenated M$_{2}$B$_{2}$, they are not superconductors (or millikelvin scale superconductors, which we neglect here), and light hydrogenation barely changes their superconducting properties. For the III-TM group, we notice only minor changes in both $\lambda$ and $\omega_{\text{log}}$ that are $\Delta\lambda$=0.068 and $\Delta\omega_{\text{log}}$=5.1meV for light hydrogenation of Sc$_{2}$B$_{2}$ and $\Delta\lambda$=0.162 and $\Delta\omega_{\text{log}}$=0.2meV for light hydrogenation of Y$_{2}$B$_{2}$. These lead to barely changes in T$_{c}$ for light hydrogenation of the III-TM group. For V-TM group, we also observed a small change in $\lambda$ and $\omega_{\text{log}}$ from non-hydrogenation to light hydrogenation, which is similar to the III-TM group. Therefore, these results in the fact that the non-hydrogenation and light hydrogenation barely become superconductors, except for V$_{2}$B$_{2}$H which could be superconducting at about 1.3K.

    For heavy hydrogenation, we still have a low electron-phonon coupling ($\lambda$) of 0.68 and 0.41 for Sc$_{2}$B$_{2}$H$_4$ and Y$_{2}$B$_{2}$H$_4$ even through we have significant increase in $\omega_{\text{log}}$. As a result, the higher $\lambda$ of Sc$_{2}$B$_{2}$H$_4$ leads to a higher T$_{c}$ = 18.7K, where the lower $\lambda$ of Sc$_{2}$B$_{2}$H$_4$ gives T$_{c}$=2.74K. However, for the V-TM group, we have significant changes in both $\lambda$ and $\omega_{\text{log}}$ from non-hydrogenation to heavy hydrogenation. The changes are $\Delta\lambda$=1.08 and $\Delta\omega_{\text{log}}$=15.73meV for V$_{2}$B$_{2}$H$_{4}$ and $\Delta\lambda$=1.11 and $\Delta\omega_{\text{log}}$=16.18meV for Nb$_{2}$B$_{2}$H$_{4}$. Due to these high values of $\lambda$ in the V-TM group, it leads to a higher T$_{c}$ for V$_{2}$B$_{2}$H$_{4}$ and Nb$_{2}$B$_{2}$H$_{4}$ compared to the III-TM group. For V$_{2}$B$_{2}$H$_{4}$, it possibly becomes superconducting at least 53.35K up to 84K. Lastly, Nb$_{2}$B$_{2}$H$_{4}$ also shows very promising superconductor of T$_c$ at least 54.40K up to 73K. All the values of T$_{c}$, $\lambda$, and $\omega_{\text{log}}$ are reported in Table~\ref{superconducting-quantities-table}.

    \section{Methods.}
	All calculations of our investigation are based on Density Functional Theory (DFT) implemented in the well-established QUANTUM ESPRESSO (QE) package \cite{giannozzi2009quantum,giannozzi2017advanced}. The norm-conserving pseudopotentials \cite{hamann2013optimized,schlipf2015optimization} and the Perdew–Burke–Ernzerhof (GGA-PBE) \cite{perdew1996generalized} are used for the exchange-correlation energy functional with wave function and charge density cutoffs of 80 Ry and 320 Ry, respectively. The vacuum thickness was set to be 30\AA\ to make sure no overlap of the wavefunction between monolayers, and the long-range Coulomb interaction was truncated along the z-axis using \cite{sohier2017density,sohier2017breakdown}. The crystal structures were optimized using the BFGS method \cite{BFGS,liu1989limited} by fully relaxing the crystal structures with a force threshold of $1.0^{-5}$ eV/\AA. The crystal structure was visualized using VESTA \cite{momma2011vesta} with layered hexagonal transition metal with $P6/mmm$ space-group symmetry. For electronic structure calculations, we used 24$\times$24$\times$1 k-point grid for self-consistent calculations to sample the reciprocal space of the Brillouin zone. The electronic density of states and Fermi surfaces were computed using the optimized tetrahedral method for the calculation of non-self-consistency, \cite{kawamura2014improved}, where the Fermi surface was visualized by XCRYSDEN \cite{kokalj2003computer}.

    To obtain the isotropic Eliashberg spectral function for and $M_2 B_2 $ and $M_2 B_2 H$, we computed the interatomic force constants (IFC) by performing the Fourier transform of the matrix elements $g_{\boldsymbol{k}+\boldsymbol{q},\boldsymbol{k}}^{\boldsymbol{q}\nu,mn}$ of the coarse 12$\times$12$\times$1 q-mesh grid calculated by using Density Functional Perturbation Theory (DFPT) \cite{baroni2001phonons} implemented in QE. From corrected IFCs, the phonon linewidth, $\gamma_{\boldsymbol{q}\nu}$, was computed,

        \begin{equation}  \label{gammaphononlinewidths}
		\gamma_{\boldsymbol{q}\nu} = 2\pi\omega_{\boldsymbol{q}\nu}\sum_{nm}\sum_{\boldsymbol{k}}|g_{\boldsymbol{k}+\boldsymbol{q},\boldsymbol{k}}^{\boldsymbol{q}\nu,mn}|^{2}\delta(\epsilon_{\boldsymbol{k}+\boldsymbol{q},m}-\epsilon_{F})\delta(\epsilon_{\boldsymbol{k},n}-\epsilon_{F})
	\end{equation}
     and the electron-phonon coupling, $\lambda_{\boldsymbol{q}\nu}$, associated with the phonon wavevector $\boldsymbol{q}$ and the phonon mode of $\mu$, as
    \begin{equation}\label{eqn:lambda_qv}
        \lambda_{\boldsymbol{q}\nu} = \frac{\gamma_{\boldsymbol{q}\nu}}{\pi N(\epsilon_{\textbf{F}})\omega_{\boldsymbol{q}\nu}^2}.
    \end{equation}
    The Eliashberg spectral function, $\alpha^{2} F(\omega)$, is obtained by
\begin{equation}
	\alpha^{2}F(\omega)=\frac{1}{2N(\epsilon_{F})}\sum_{\boldsymbol{q}\nu}\frac{\gamma_{\boldsymbol{q}\nu}}{\omega_{\boldsymbol{q}\nu}} \delta(\omega-\omega_{\boldsymbol{q}\nu}).
\end{equation}

    To investigate $M_2 B_2 H_4$, we employ the electron-phonon Wannier-Fourier interpolation method \cite{qiao2023automated,giustino2017electron,giustino2007electron} within the EPW package \cite{noffsinger2010epw,ponce2016epw} to accurately compute the superconducting properties, specifically the electron-phonon coupling ($\lambda$) and critical temperature (T$_{c}$). This methodology also allows for the detailed analysis of the anisotropic Migdal-Eliashberg theory \cite{frohlich1950theory,migdal1958interaction,eliashberg1960interactions,nambu1960quasi,margine2013anisotropic} by solving the two coupled nonlinear anisotropic Migdal-Eliashberg equations,
\begin{align}
    Z_{nk}(i\omega_{j}) &= 1 + \frac{\pi T}{N(\varepsilon_{F})} \sum_{mk'j'} \frac{\omega_{j'}}{\sqrt{\omega_{j'}^2 + \Delta_{mk'}^2(i\omega_{j'})}}, \\
    Z_{nk}(i\omega_{j})\Delta_{mk}(i\omega_{j}) &= \frac{\pi T}{N(\varepsilon_{F})} \sum_{mk'j'} \frac{\Delta_{mk'}(i\omega_{j'})}{\sqrt{\omega_{j'}^2 + \Delta_{mk'}^2(i\omega_{j'})}} \left[ \lambda(nk, mk', \omega_j - \omega_{j'}) - \mu^* \right] \delta(\epsilon_{mk'} - \varepsilon_F),
\end{align}
    self-consistently along the imaginary axis at the fermion Matsubara frequencies $\omega_j = (2j+1)\pi T$. For these calculations, we use the QUANTUM ESPRESSO package with the same computational settings as previously mentioned, followed by Wannier-Fourier interpolation to k- and q-point grids of $120 \times 120 \times 1$ and $60 \times 60 \times 1$, respectively. Employing dense grids ensures the convergence of $\lambda$ values, as indicated by the stability of the corresponding $\alpha^2F(\omega)$ and $\lambda(\omega)$ even as the k- and q-point grid densities increase. The Fermi surface thickness was set to 0.55 eV, with a Matsubara frequency cutoff at 1.35 eV. Dirac $\delta$ functions were broadened using a Gaussian function with widths of 0.1 eV for electrons and 0.5 meV for phonons when analysing $V_2 B_2 H_4$, and $Nb_2 B_2 H_4$. Due to maximun phonon frequency, the numerical parameters for $Sc_2 B_2 H_4$, and $Y_2 B_2 H_4$ were set accordingly. The Fermi surface thickness was set to 0.68 eV, with a Matsubara frequency cutoff at 1.7 eV while we used the same Dirac $\delta$ functions for electrons and for phonons. Lastly, the Morel-Anderson pseudopotential was set to $\mu^* = 0.1$ for practical purposes.

The superconducting transition temperature (T$_{C}$) was determined using the semi-empirical Allen-Dynes formula \cite{allen1975transition}:

\begin{equation}
    T_{c} = f_1 f_2 \frac{\omega_{\text{log}}}{1.20} \exp\left(-\frac{1.04(1+\lambda)}{\lambda - \mu^*(1 + 0.62\lambda)}\right)
\end{equation}

where the electron-phonon coupling constant $\lambda$ is derived from the Eliashberg spectral function:

\begin{equation}
    \lambda = 2 \int^{\omega}_{0} d\Omega \left(\frac{\alpha^2F(\Omega)}{\Omega}\right),
\end{equation}

and the logarithmic average phonon frequency is calculated as:

\begin{equation}
    \omega_{\text{log}} = \exp\left(\frac{2}{\lambda}\int^{\infty}_{0} d\Omega \log(\Omega) \left(\frac{\alpha^2F(\Omega)}{\Omega}\right)\right).
\end{equation}

The correction factors $f_1$ and $f_2$ are given by:

\begin{equation}
    f_1 f_2 = \left(1 + \left(\frac{\lambda}{2.46(1 + 3.8\mu^*)}\right)\right)^{1/3} \times \left(1 + \frac{\lambda^2 \left(\frac{\omega_2}{\omega_{\text{log}}} - 1\right)}{\lambda^2 + 3.31(1 + 6.3\mu^*)^2}\right).
\end{equation}

This $f_1 f_2$ correction factor is applied when the electron-phonon coupling constant $\lambda$ exceeds 1.0. The mean-square phonon frequency ($\omega_2$) is given by
\begin{equation}
    \omega_2 = \sqrt{\frac{2}{\lambda}\int_{0}^{\omega_{\text{max}}}\alpha^2 F(\omega)\omega d\omega}.
\end{equation}


	\section{CONCLUSION.}
	In this work, we systematically investigate 2D transition metal borides M$_{2}$B$_{2}$ (M = Sc, Y, V, and Nb) with different levels of hydrogenation. The non-hydrogenation M$_{2}$B$_{2}$ has been suggested to have dynamical and thermal stability. For M$_{2}$B$_{2}$, we again confirm these suggestions with negative formation energies and non-negative phonon dispersion based on DFT computations. For light hydrogenation M$_{2}$B$_{2}$H and heavy hydrogenation M$_{2}$B$_{2}$H$_{4}$, we also found that they are dynamically stable with formation energies even lower than the 2D pristine M$_{2}$B$_{2}$. For electronic band structures, we found that the light hydrogenation M$_{2}$B$_{2}$H can be described as a perturbation of the 2D pristine M$_{2}$B$_{2}$. This leads to minor changes in the electronic bands and in the electronic density of states. A major modification of the electronic bands occurs when applying heavy hydrogenation M$_{2}$B$_{2}$H$_{4}$. This results in many changes from M$_{2}$B$_{2}$ and M$_{2}$B$_{2}$H to M$_{2}$B$_{2}$H$_{4}$. As a result, these results suggest that the disrupted evolution of the electronic band topology of Fermi surfaces can be slightly changed and heavily disrupted because of the amount of hydrogenation. For the III-TM group (Sc, Y), we found that EDOS decreases significantly at the Fermi level. The reduction in EDOS occurs due to the decrease of d-orbital electrons as a result of the bonding of additional hydrogen atoms, especially for Y$_{2}$B$_{2}$H$_{4}$ in which the d-orbital electrons no longer completely dominate at the Fermi level. On the other hand, the V-TM group (Sc, Y) shows a higher EDOS compared to the III-TM group, and even higher for V$_{2}$B$_{2}$H$_{4}$ because of a significant van Hove singularity located almost at the Fermi level. This would favor $\lambda$ for the heavy hydrogenation of the V-TM group rather than the heavy hydrogenation of the III-TM group. For phonons, the phonon dispersion of light hydrogenation is similar to that of non-hydrogenation, with three additional optical modes of hydrogen atom dominating the highest phonon spectrum. For the heavy hydrogenation M$_{2}$B$_{2}$H$_{4}$, the phonon dispersion is completely disturbed, and the new phonon spectrum eventually leads to new superconducting properties. Even through, $\omega_{\text{log}}$ is higher for the heavy hydrogenation of the III-TM group than that of the V-TM group, $\lambda$ is higher for the heavy hydrogenation of V-TM group than that of III-TM group. As a result, Nb$_{2}$B$_{2}$H$_{4}$ shows the most promising superconductor with T$_{c}$ of at least 40.1K where T$_{c}$ of V$_{2}$B$_{2}$H$_{4}$, Sc$_{2}$B$_{2}$H$_{4}$, and Y$_{2}$B$_{2}$H$_{4}$ are at least 33.4, 18.3, and 4.3K, respectively.

\backmatter

\bmhead{Acknowledgments}

This research project is supported by the Second Century Fund (C2F), Chulalongkorn University. The authors acknowledge the National Science and Technology Development Agency, National e-Science Infrastructure Consortium, Chulalongkorn University and the Chulalongkorn Academic Advancement into Its 2nd Century Project (Thailand) for providing computing infrastructure that has contributed to the research results reported within this paper. URL:www.e-science.in.th.  GJA acknowledges funding from the ERC project Hecate. The authors acknowledge NSTDA Supercomputer Center (ThaiSC) for 
providing LANTA computing resources for this work. This also work used the Cirrus UK National Tier-2 HPC Service at EPCC (http://www.cirrus.ac.uk) funded by the University of Edinburgh and EPSRC (EP/P020267/1).


\bibliography{sn-bibliography}

\section*{COMPETING INTERESTS}

The authors declare no competing financial or non-financial interests.

\section*{DATA AVAILABILITY}

Data will be available from the authors on request.

\section*{AUTHOR CONTRIBUTIONS}

Jakkapat Seeyangnok performed structural, electronic, phonon and superconductivity calculations, analysed the results, and wrote the first draft manuscript.
Udomsilp Pinsook coordinated the project, analysed the results, and wrote the manuscript. 
Graeme Ackland analysed the results, supervised the project, and wrote the final manuscript. 
All authors have approved the final manuscript.

\end{document}